\documentclass[
 reprint,
 amsmath,amssymb,
 aps,
 pra,
]{revtex4-2}

\usepackage{graphicx}
\usepackage{dcolumn}
\usepackage{bm}
\usepackage[colorlinks,urlcolor=blue,linkcolor=blue,citecolor=blue]{hyperref}
\usepackage[capitalise]{cleveref}
\newcommand{\intd}{\,\mathrm{d}}

\newcommand{\tee}{\intercal}

\begin{document}

\title{Atom Interferometer Phase Shear and Spacetime Sectional Curvature}

\author{Hunter Swan}
 \email{orswan@stanford.edu}
\author{Jason M. Hogan}
 \email{hogan@stanford.edu}
\affiliation{
 Department of Physics, Stanford University, Stanford, California 94305, USA
}
\date{\today}

\begin{abstract}
Atom interferometry is a natural laboratory for precision tests of general relativity, but there are no known simple relationships between atom interferometer phase and geometric properties of spacetime.  Here we show that a different atom interferometer observable, the phase shear, can be expressed directly as the integrated sectional curvature over a spacetime surface enclosed by the interferometer arms and final beamsplitter, which follows from a generalized Gauss-Bonnet theorem. This interface of quantum mechanics, relativity, and geometry is manifestly coordinate-free and representation-free and can provide a useful computational tool for predicting atom interferometer properties in arbitrary background spacetimes. 
\end{abstract}

\maketitle

\section{Introduction}
From the first demonstration of a light-pulse atom interferometer (AI), this platform has been understood to be an excellent technique for precision measurement of gravity \cite{kasevich1992measurement}.  Atom interferometry now spans a wide range of gravimetric applications such as geodesy \cite{peters2001high,stockton2011absolute}, gravitational wave detection \cite{Dimopoulos2009,Graham2013,AGIS2008,MAGIS2021}, and weak equivalence principle tests \cite{Bonnin2013,Schlippert2014,Tarallo2014,Zhou2015,Asenbaum2020}.  Although many formalisms exist for describing and predicting properties of AIs \cite{GRAI,Badurina2025,Werner2024,Roura2025,Roura2020}, the relationship between these properties and intrinsic geometric features of spacetime is often opaque. This is due partly to the confounding influence of non-gravitational effects and partly to the algebraic difficulty of modeling typical AI geometries. 

The main observable in an AI is the phase $\Delta \phi$.  Detailed fully relativistic calculations in a background Schwarzschild spacetime \cite{GRAI} show that spacetime geometry does contribute significantly to observed phases $\Delta\phi$, an effect which has also been observed experimentally \cite{asenbaum2017phase}.  However, these geometric contributions are mixed together with phase shifts arising from a variety of other parasitic effects, such as laser noise.  Moreover, typical expressions for $\Delta\phi$ invoke frame dependent quantities such as the local gravitational acceleration $g$, which have no intrinsic geometric meaning.  Significant effort has been devoted towards clarifying the geometric content of AI phases.  On the one hand, specialized AI sequences have been designed to suppress parasitic phase shifts, such as the gradiometer configuration of \cite{Graham2013} which eliminates laser frequency noise via a differential measurement between two interferometers.  On the other hand, investigations as in \cite{overstreet2021physically} have examined which aspects of AI phase shift calculation methods are physically meaningful, i.e.\ independent of arbitrary choices of coordinates or decomposition of the AI phase.  It is therefore of interest to understand which properties of an AI are related to intrinsic properties of spacetime. 

In this work, we show that in contrast to $\Delta\phi$, the \textit{phase shear} of an AI has a direct geometric meaning related to integrated sectional curvature over a spacetime surface bounded by the interferometer arms and final beamsplitter (FBS).  Phase shear arises due to a momentum difference between the interferometer arms after the FBS, i.e.\ a failure to close in momentum space, and manifests as a spatially periodic modulation of the interferometer output ports.  Phase shear due to a gravity gradient has been measured experimentally \cite{overstreet2018effective} and is an important effect for AI weak equivalence principle tests \cite{roura2017circumventing}. 

Our basic result states that if an AI is designed such that in flat space the momentum difference $\Delta p$ between the two wavepackets in a given output port would vanish, then applying the same laser pulse sequence to an AI in curved space will lead to a momentum difference 
\begin{equation}
\label{eq:main}
\Delta p = -m c \int_M K \intd{A} + \mathcal{O}\left(m\bar v \bar K \bar A + mc \bar K^2 \bar A^2\right),
\end{equation}
where $m$ is the atom mass, $c$ the speed of light, $M$ a spacetime surface bounded by the AI arms and FBS, $K$ the sectional curvature on $M$, and $\intd{A}$ the area element on $M$. The notation $\mathcal{O}(x)$ indicates additional terms of typical magnitude $x$ or smaller, and $\bar v, \bar K, \bar A$ are characteristic scales for the atomic recoil velocity, spacetime sectional curvature, and spacetime area, respectively. The magnitude of $m\bar v \bar K \bar A$ is smaller than the first term in \cref{eq:main} by $\bar v/c$, which is typically $\sim10^{-11}$.  The relative magnitude of $mc\bar K^2 \bar A^2$ depends on the interferometer sequence, but for typical experiments it is much smaller than $m\bar v \bar K \bar A$. 
We will give explicit expressions for these correction terms. We emphasize \cref{eq:main} applies equally to multi-photon atom optics (such as Bragg or Raman transitions) or single-photon atom optics \cite{Graham2013}.  

The momentum difference $\Delta p$ induces a modulation on the AI output port with wavelength $\lambda = \frac{h}{|\Delta p|}$, where $h$ is Planck's constant.  The observability of this effect depends roughly on whether this modulation wavelength is smaller than the AI atom cloud size, which is typically a few millimeters.  Note that phase shear can also arise from non-gravitational effects such as frequency detuning \cite{roura2017circumventing}, wavepacket curvature \cite{sugarbaker2013enhanced}, or transverse phase gradients of atom optics pulses \cite{roura2017circumventing,sugarbaker2013enhanced}.  Phase shear due to transverse phase gradients differs in a qualitative way from that of gravitational origin by causing spatial modulation transverse rather than parallel to the interferometer baseline.  On the other hand, frequency detuning of atom optics pulses has an interesting interplay with gravitationally induced phase shear, and we will discuss variants of \cref{eq:main} relevant to such cases.  We will not consider other sources of phase shear besides gravitation and frequency detuning here, assuming all wavepackets are collimated throughout the interferometer.

The hypothesis that an AI pulse sequence would result in no phase shear if executed in flat space serves as a simple mathematical criterion for isolating the effect of gravity and excluding other sources of phase shear.  It is also compatible with the manner in which many typical AI experiments are performed, where e.g.\ the laser frequency may be detuned to compensate for the fact that the laser is not in free fall, but not in such a way as to deliberately induce shear.  We refer to this hypothesis as the closure condition, and we emphasize that this is not an essential limitation of our method, but rather a simple criterion which allows us to make statements about a broad class of AI sequences.  There are interesting examples of AI sequences which violate the closure condition, such as those of \cite{roura2017circumventing,overstreet2018effective} where phase shear is eliminated by a judicious frequency offset of the central mirror pulse.  We will see in this case the analog of \cref{eq:main} relates spacetime curvature to this frequency offset rather than to phase shear.

In derivations below we adopt natural units $c=\hbar=1$.  We will reinsert dimensional constants when we consider specific examples. 

\section{Calculation of Phase Shear}
Relating AI phase shear to spacetime curvature requires three main insights, which may be described as the closure condition, Gauss-Bonnet, and redshift steps.

The first step, the closure condition, accounts for the contribution to phase shear due to the atom optics laser pulses.  It is possible in some situations (namely, when using two-photon atom optics) to set AI phase shear to any desired value by detuning one or more of the laser pulses, and the formalism must thus take into account this induced shear.  To be able to make general statements that allow for a wide range of AI geometries, such as large momentum transfer (LMT) or resonant pulse sequences \cite{wang2024robust}, we impose a condition that whatever pulse sequence is chosen, when executed in a corresponding flat spacetime with lasers following trajectories of equal geodesic curvature, it should result in zero phase shear. 

The second step applies a Gauss-Bonnet formula to evaluate the final momentum separation of the AI arms in curved space, using the closure condition.  This employs a formulation of the Gauss-Bonnet formula for Lorentzian manifolds which allows for null geodesic boundary segments.  It is found that the final momentum difference is given by a spacetime curvature integral plus additional terms that depend only on the redshift of atom optics pulses. 

The third step, redshift, calculates the additional terms in the previous step.  These terms depend on the nature of the atom optics pulses which are employed, and in particular on whether two-photon or single-photon processes are used.  In all cases we show that these terms are typically small compared to the leading term in \cref{eq:main}.  For two-photon processes, evaluating these redshift terms involves a second application of the Gauss-Bonnet formula to auxiliary surfaces bounded by the laser and atom trajectories.  For single-photon processes, under certain assumptions of semiclassical dynamics it can be shown the correction terms vanish completely. 

\subsection{Lorentzian Angle and Gauss-Bonnet}
Our main technical tool is a generalized version of the Gauss-Bonnet (GB) formula.  Many authors \cite{Helzer,BirmanNomizu, Dzan, Law} have introduced Lorentzian versions of GB, though most exclude null boundary segments.  We use a formalism similar to \cite{BirmanNomizu, Law}, but modified to allow null geodesics, as described in Appendix A.  This relativistic GB formula states that, given a piecewise smooth curve $\gamma$ bounding a 1+1 dimensional Lorentzian surface $M$, under appropriate technical hypotheses we have 
\begin{equation}
\label{eq:GB}
\int_M K\intd{A} + \oint_\gamma k_g \intd{s} + \sum_j\alpha_j = 0,
\end{equation}
where $K$ is sectional curvature, $k_g$ is geodesic curvature along $\gamma$, $\intd{A}$ is the area element on $M$, $\intd{s}$ is differential line element along $\gamma$, and $\alpha_j$ is an ``angle'' (to be defined below) associated to each non-smooth point $P_j$ on $\gamma$. 

The sectional curvature and geodesic curvature terms of \cref{eq:GB} are identical to the usual (Riemannian) GB formula \cite{hicks},
but care is needed to define a notion of angle on a Lorentzian manifold.  Given two time-like future directed unit vectors $S,\, T$ at a point $P\in M$, we define the angle $\alpha$ from $S$ to $T$ to be the boost angle (i.e.\ rapidity) of a Lorentz transformation $L(\alpha)$ acting on the tangent space $\mathbf{T}_{\! P} M$ which maps $S$ to $T$.  Recall that with respect to an orthonormal basis for $\mathbf{T}_{\! P} M$, $L(\alpha)$ takes the form 
\begin{equation}
\label{eq:LT}
L(\alpha) = \begin{pmatrix} \cosh(\alpha) & \sinh(\alpha) \\ \sinh(\alpha) & \cosh(\alpha) \end{pmatrix}.
\end{equation}
For an arbitrary non-null vector $S \in \mathbf{T}_{\! P} M$, let $\widehat S$ denote the unique time-like future directed unit vector in $\mathbf{T}_{\! P} M$ which is either parallel to $S$ (if $S$ is time-like) or orthogonal to $S$ (if $S$ is space-like).  Then given any two such non-null vectors $S,\,T \in \mathbf{T}_{\! P} M$, the angle from $S$ to $T$ is defined to be the angle from $\widehat{S}$ to $\widehat{T}$. Angles so defined have an additivity property: The angle from $S$ to $T$ plus the angle from $T$ to $U$ equals the angle from $S$ to $U$.

\begin{figure}[t]
\centering
    \includegraphics[width=0.9\linewidth]{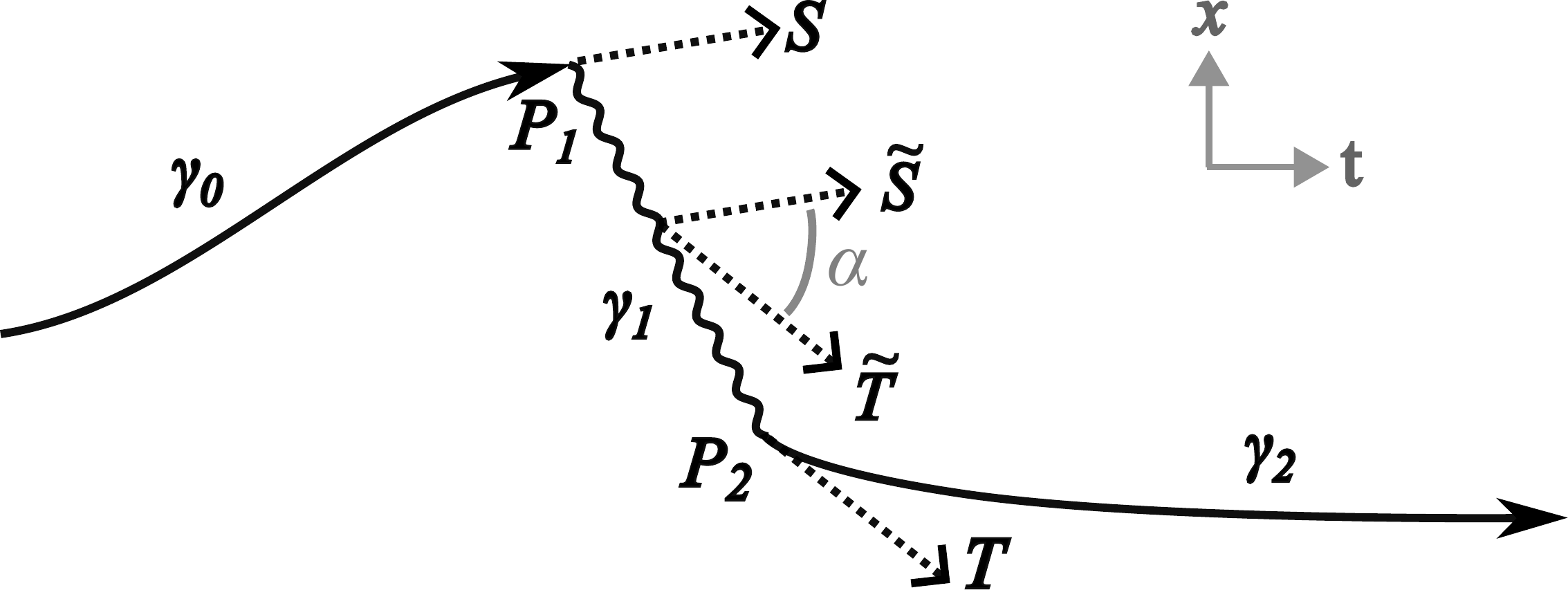}
    \caption{Defining angle $\alpha$ at a null geodesic $\gamma_1$ by parallel transport of tangent vectors. The wavy line is a null geodesic, solid lines are time-like curves, and dotted lines are tangent vectors.  $\alpha$ is defined to be the angle from $\widetilde S$ to $\widetilde T$.}
\label{fig:parallel-transport} 
\end{figure}

Given two time-like vectors $S$ and $T$, the angle from $S$ to $T$ can be conveniently evaluated in the rest frame of $S$, i.e.\ the local Lorentz frame such that $\widehat S$ has components $(1,0)^{\tee}$.  Let $T$ have components $(T_t,T_x)^{\tee}$ in this frame.  Then the angle from $S$ to $T$ is given by $\tanh^{-1}(T_x/T_t)$. 

Finally, we consider angles associated to null geodesics (see \cref{fig:parallel-transport}).  Suppose that null geodesic segment $\gamma_1$ intersects two non-null curves $\gamma_0$ and $\gamma_2$ at points $P_1,\,P_2$ respectively, and define a single angle $\alpha$ for these two points as follows.  Take the tangent vector $S$ to $\gamma_0$ at $P_1$ and the tangent vector $T$ to $\gamma_2$ at $P_2$ and parallel transport both along the null geodesic $\gamma_1$ to any common point.  Write the parallel transported vectors $\widetilde S,\, \widetilde T$ and define $\alpha$ to be the angle from $\widetilde S$ to $\widetilde T$.  This definition is independent of the specific point along $\gamma_1$ chosen.  For convenience, in the context of \cref{eq:GB} we define the angle of the initial endpoint of a null geodesic to be this $\alpha$ and the angle of the final endpoint to be zero.  

More general notions of Lorentzian angle are possible \cite{Helzer}, but that above is convenient and sufficient for our purpose. With these definitions, \cref{eq:GB} holds provided all null segments of $\gamma$ are geodesic and no two null segments are adjacent.  See Appendix A for more details.

\subsection{Recoil Boosts}

Before applying \cref{eq:GB} to an AI, we recall some standard relationships between boost angle and momentum. Consider a particle $\Pi$ with mass $m$ and unit tangent vector $T$, and let $S$ be any other time-like future directed vector at the same point. Let $\alpha$ be the angle from $S$ to $T$. 
In the rest frame of $S$, the velocity of $\Pi$ is $\tanh(\alpha)$ and the momentum of $\Pi$ has components $m(\cosh(\alpha),\sinh(\alpha))^{\tee}$, so that the spatial component is
\begin{equation} 
\label{eq:deltap}
p = m \sinh(\alpha) = m\alpha + \mathcal{O}\left(m\alpha^3\right).
\end{equation}

If $\Pi$ absorbs a photon with momentum $(\nu,\nu)^{\tee}$ in the rest frame of $T$, its final momentum is $(m+\nu,\nu)^{\tee}$, and its recoil velocity $v$ is $\nu/(m+\nu)$, which has typical magnitude $\sim 2\times 10^{-11}$ for e.g.\ the $698\text{ nm}$ clock transition in $^{87}\text{Sr}$. The angle from initial to final momentum of $\Pi$ is 
\begin{equation}
\label{eq:recoil}
\beta = \tanh^{-1}\left(\frac{\nu}{m+\nu}\right) = \frac{\nu}{m+\nu} + \mathcal{O}\left(\bar{v}^3\right).
\end{equation} 
Similarly, if $\Pi$ then emits an identical photon, it experiences a boost $-\beta$.

An important subtlety arises when an atom interacts with an atom optics pulse which is not perfectly on resonance.  In this case, the effective photon momentum transferred to the atom through absorption or emission may differ from the nominal momentum of a photon at the pulse carrier frequency, due to the non-zero pulse bandwidth.  As described in \cite{swan2026exact}, we follow the frequency selectivity prescription for computing the boost imparted to an atom in such a case.  For a single-photon transition, this implies that the boost imparted by an off-resonant atom optic always coincides with that of a resonant pulse. 

For an atom optic utilizing a two-photon transition, 
we consider the case where the lower laser is a passive laser which is always on, while the upper laser is active, such that its pulse envelope determines the spacetime position of the atom-light interaction.  As shown in \cite{swan2026exact}, assuming the frequency selectivity prescription and short pulse limit, the boost imparted to the atom when absorbing from the lower laser and emitting into the upper laser can be expressed in terms of the frequency $\nu_-$ of the lower laser in the atom's rest frame (i.e.\ the rest frame of $T$) as 
\begin{equation}
\label{eq:recoil_two_photon}
\beta = \ln\left(\frac{m + 2\nu_-}{m+\Delta \omega_a}\right) = \frac{2\nu_- - \Delta\omega_a}{m} + \mathcal{O}\left( \bar{v}^2\right),
\end{equation}
where $\Delta \omega_a$ is the hyperfine splitting of the atom's ground state levels for a Raman transition, or for a Bragg transition $\Delta\omega_a =0$.
If the laser frequency is given as $\nu_-$ in the rest frame of $S$ rather than $T$, then to compute the boost imparted to the atom we first compute the photon momentum in the atom's rest frame via
\begin{equation}
\label{eq:freq_shift}
L(-\alpha) \begin{pmatrix} \nu \\ \nu\end{pmatrix} = \begin{pmatrix} e^{-\alpha}\nu \\ e^{-\alpha}\nu\end{pmatrix}
\end{equation}
and then substitute $e^{-\alpha}\nu_-$ for $\nu_-$ in \cref{eq:recoil_two_photon}. 

\subsection{Atom Interferometer Phase Shear}

\begin{figure}[t]
\centering
\includegraphics[width=\linewidth]{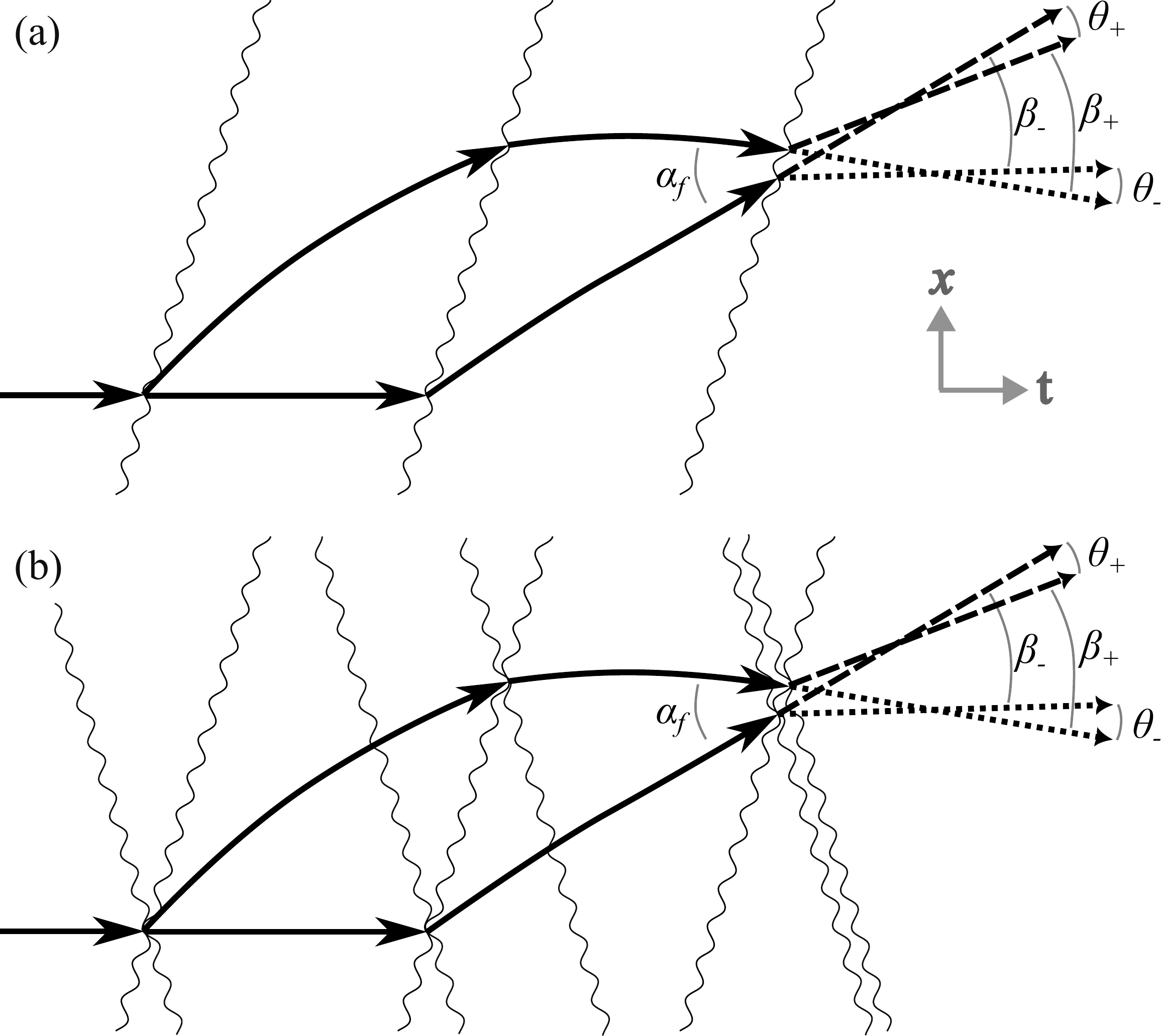}
\caption{\label{fig:AI} Typical Mach-Zehnder atom interferometers. Thick solid lines are geodesic atom trajectories. Dashed lines are output port atom wavepackets. Wavy lines are atom optics null geodesics. (a) Single-photon interferometer. (b) Two-photon (e.g.\ Raman or Bragg) interferometer.}
\end{figure}

We now apply \cref{eq:GB} to an AI, as shown schematically in \cref{fig:AI}. We follow the covariant semiclassical formalism of \cite{GRAI} in describing the AI.  The upper and lower output ports of the AI both consist of two wavepackets arising from the two AI arms, and the phase shear of the upper (lower) output port can be characterized by a boost angle $\theta_+$ ($\theta_-$) between these two wavepackets along the final beamsplitter null geodesic. We refer to $\theta_\pm$ as a \textit{shear angle}, and it determines the momentum difference $\Delta p$ between the output port wavepackets via \cref{eq:deltap}. Note that generically $\theta_+ \neq \theta_-$, leading to (slightly) different phase shear between the two output ports. 

We set $\gamma$ in \cref{eq:GB} to consist of the lower AI arm, followed by the FBS null geodesic segment connecting the two arms, followed by the upper arm traversed backwards. Since all boundary segments are geodesic, geodesic curvature is uniformly zero on $\gamma$, and \cref{eq:GB} relates angles on $\gamma$ (i.e.\ the holonomy of $\gamma$) to the enclosed sectional curvature. Let $\alpha_f$ denote the angle across the FBS, and let $-\beta_-$ ($\beta_+$) be the boost imparted to the lower (upper) arm by the FBS. Note that $\beta_\pm$ is not an angle of $\gamma$, but is needed to compute the shear angle $\theta_\pm$, which is given by 
\begin{equation}
\theta_\pm = \alpha_f + \beta_\pm.
\end{equation}

We consider the typical case where the atom optics pulses are designed such that the shear angle would vanish for an AI in flat space (the closure condition), and examine what the shear angle must be for an AI in curved space with atom optics pulses of the same order, direction, and frequency at point of emission, and with lasers following trajectories of equal geodesic curvature.  Since we need to refer to analogous quantities in both flat space and curved space AIs, we adopt the convention that unprimed symbols such as $\alpha_f, \theta_+$ refer to a curved space AI, while primed quantities $\alpha_f', \theta_+'$ refer to analogous quantities in a flat space AI, and a differential quantity like $\Delta\alpha_f$ will denote $\alpha_f - \alpha_f'$.  The closure condition thus takes the form $\theta_\pm' = 0$. 

In the flat space AI, \cref{eq:GB} becomes 
\begin{equation}
\label{eq:FSAI}
\sum_j \alpha_j' = 0
\end{equation}
since $K'=0$.  (Recall $k_g'=0$ because all segments are geodesic.) In curved space, \cref{eq:GB,eq:FSAI} give
\begin{align}
0 & = \int_{M} K \intd{A} + \sum_j \alpha_j = \int_{M} K \intd{A} + \sum_j \Delta\alpha_j.
\end{align}
Solving for $\Delta\alpha_f$, 
\begin{equation}
\Delta\alpha_f = -\int_{M} K \intd{A} - \sum_{j\neq f} \Delta\alpha_j,
\end{equation}
which together with $\theta_\pm'=0$ gives the shear angle
\begin{align}
\label{eq:shear}
\theta_\pm & = \alpha_f + \beta_\pm = \Delta\alpha_f + \Delta\beta_\pm \nonumber \\
& = -\int_{M} K \intd{A} - \sum_{j\neq f} \Delta\alpha_j + \Delta\beta_\pm.
\end{align}
It remains to compute $\Delta\alpha_j$ for $j\neq f$ and $\Delta\beta_\pm$.  Note that $\alpha_j$ ($j\neq f$) and $\beta_\pm$ depend only on the frequencies of the atom optic photons at the points of interaction, as in \cref{eq:recoil,eq:recoil_two_photon}, unlike $\alpha_f$, which depends on the relative motion of the two AI arms at the FBS.  Thus $\Delta\beta_\pm$, $\Delta\alpha_j$ depend only on the change in photon frequency due to spacetime curvature, i.e.\ gravitational redshift.  We thus refer to such quantities as redshift corrections to phase shear, and the objective of the remainder of our calculation is to show that these terms are small compared to the first term in \cref{eq:shear}.

The character of redshift corrections depends on whether single-photon or two-photon transitions are employed.  In the single-photon case, under the frequency selectivity prescription discussed above, the boost angles imparted by atom optics are unaffected by the pulse carrier frequency.  
Thus in this case, the redshift of an atom optics pulse has no effect whatsoever on the recoil boost, and so all recoil corrections are identically zero.  From \cref{eq:deltap,eq:shear} we thus find that the momentum difference $\Delta p$ (in the rest frame of the output port wavepacket from the lower arm) is given by 
\begin{align}
\label{eq_freq_selectivity_result}
\Delta p & = -m \sinh\left(\int_M K \intd{A}\right) \\
& = -m \int_M K \intd{A} + \mathcal{O}\left( m\bar K^3 \bar A^3 \right)
\end{align}
For the single-photon case we thus obtain a refined version of \cref{eq:main} which is exact to all orders within the semiclassical formalism of \cite{GRAI}.  

We next sketch the calculation of redshift corrections for two-photon transitions. For simplicity, we consider the Bragg case $\Delta\omega_a=0$ in \cref{eq:recoil_two_photon}.  
To evaluate $\Delta \alpha_j$ in this case, we apply \cref{eq:GB} to a loop bounded by the trajectories of the atom, the lower (passive) laser, and certain atom optics. For clarity, we work only to leading order in all small parameters, but this is not an essential limitation. We allow the laser to follow a non-geodesic trajectory (e.g.\ fixed to Earth's surface), but suppose that its geodesic curvature is identical in the curved and flat space AIs, i.e.\ $k_g = k_g'$, and that the laser's proper time intervals between pulse emissions also coincide between the two cases.  We further suppose for simplicity that the AI initial conditions are such that the angle $\phi_0$ across the initial beamsplitter (IBS) from the laser to lower AI arm is identical between curved and flat space, i.e.\ $\phi_0 = \phi'_0$. 

\begin{figure}[t]
\centering
\includegraphics[width=0.9\linewidth]{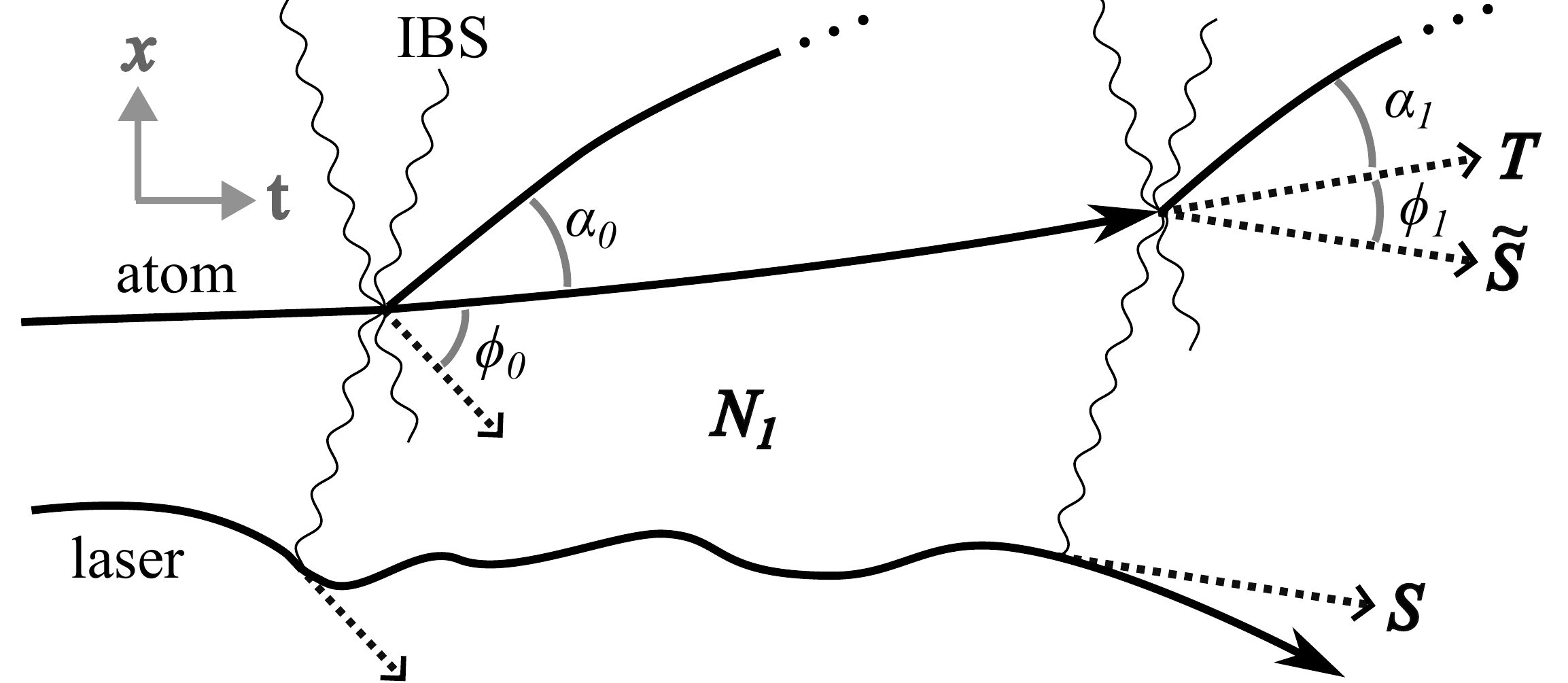}
\caption{\label{fig:redshift_square} Spacetime loop for computing the redshift of a two-photon atom optic light pulse. The segments shown represent only that portion of the atom interferometer relevant for calculating redshift effects on the initial atom optics.  The laser trajectory is shown with wiggles to emphasize that it may follow a non-geodesic trajectory.}
\end{figure}

Let $\alpha_1$ denote the boost imparted to the lower AI arm by the first atom optic after the IBS, and let $\phi_1$ denote the angle from laser to incoming atom across this atom optic null geodesic, as in \cref{fig:redshift_square}.  Letting $N_1$ denote the surface bounded by this atom optic along with the laser, atom, and IBS, \cref{eq:GB} gives 
\begin{align}
\phi_1' & = \phi_0 - \int_{\text{laser}}k_g\intd{s} \label{eq:flat_redshift_GB},\\
\phi_1 & = \phi_0 - \int_{\text{laser}}k_g\intd{s} - \int_{N_1} K \intd{A} .\label{eq:curved_redshift_GB}
\end{align}

Let $(\nu_-,\nu_-)^{\tee}$ be the momentum of the photon inducing $\alpha_1$ in the rest frame of the laser.  From \cref{eq:freq_shift}, the photon momentum in the atom's rest frame is $(e^{-\phi_1}\nu_-,e^{-\phi_1}\nu_-)^{\tee}$.
From \cref{eq:recoil_two_photon}, the atom's boost is 
\begin{align}
\alpha_1' & = \ln\left(1 + \frac{2e^{-\phi_1'}\nu_-}{m}\right) \\
\alpha_1 & = \ln\left(1 + \frac{2e^{\int_{N_1} K \intd{A}}e^{-\phi_1'}\nu_-}{m}\right) \label{eq:curved_redshift}
\end{align}
Finally, expanding to leading order in $\int_{N_1} K \intd{A}$ we find
\begin{align}
\label{eq:delta_alpha_1}
\Delta\alpha_1 & = \alpha_1' \int_{N_1} K \intd{A} + \mathcal{O}\left(\bar v^2 \bar K \bar A + \bar v \bar K^2 \bar A^2\right).
\end{align}
In all examples below it will be true that $\left|\int_{N_1} K \intd{A} \right| \ll 1$, justifying the expansion. We see that $\Delta\alpha_1$ has the same form as the leading term $\int_M K \intd{A}$ of \cref{eq:shear}, integrated over a possibly larger surface $N_1$ but suppressed by $\alpha_1'$, which is $\mathcal{O}(\bar{v})$.  Subsequent terms $\Delta\alpha_j$ and $\Delta\beta_\pm$ may be computed similarly, inductively taking into account the previous angles, with the same leading order result 
\begin{align}
\label{eq:delta_alpha_j}
\Delta\alpha_j & = \alpha_j' \int_{N_j} K \intd{A} + \mathcal{O}\left(\bar v^2 \bar K \bar A + \bar v \bar K^2 \bar A^2\right),
\end{align}
$N_j$ now being a surface bounded by the relevant atom optic, IBS, atom and laser. Together with \cref{eq:deltap,eq:shear}, this yields a refined version of our main result \cref{eq:main} for the case of two-photon atom optics,
\begin{align}
\Delta p_\pm =\; & -m \int_M K \intd{A} \nonumber \\
& - m\sum_{j\neq f} \alpha_j' \int_{N_j} K \intd{A} + m \beta_\pm'\int_{N_\pm} \! K \intd{A} \nonumber \\
& + \mathcal{O}\left(m\bar K^3 \bar A^3 + m\bar v^2 \bar K \bar A + m\bar v \bar K^2 \bar A^2\right), 
\end{align}
where the shear may now differ slightly between the two output ports.  We will not analyze the new correction terms in detail, but note that the alternating sign of $\alpha_j'$ for different $j$ will lead to significant cancellation.  This concludes the proof of \cref{eq:main}.

We next evaluate a few variations of this argument.  Firstly, we have considered a comparison to an AI in flat space, but one could similarly compare to any reference background spacetime.  If atom optics are designed to yield zero phase shear in such a reference spacetime, then observation of phase shear in an unknown spacetime can reveal deviation of the real sectional curvature from that of the reference. Secondly, in the case of multi-loop or resonant AIs in which the arms cross \cite{graham2016resonant}, one must account for the orientation of spacetime area, which varies with the handedness of the boundary, but the above argument is otherwise unchanged. Thirdly, in the procedure of \cite{overstreet2018effective,roura2017circumventing}, the central mirror pulse of a two-photon atom interferometer is detuned so that phase shear vanishes. Here $\theta_\pm = 0$, and the boost angles $\alpha_{\pi\pm}$ of the upper ($+$) and lower ($-$) interferometer arms across the central mirror pulse are adjusted between flat space and curved space by an amount
\begin{equation}
\label{eq:FSSG}
\Delta \alpha_{\pi\pm} = \pm\frac{1}{2}\int_{M} K \intd{A} + \mathcal{O}(\bar v \bar K \bar A)
\end{equation}
so that \cref{eq:GB} remains satisfied.  From \cref{eq:recoil_two_photon,eq:freq_shift}, the frequency shift $\Delta\nu_-$ which realizes this is $\frac{1}{2}m\Delta\alpha_{\pi-}$ to leading order. Thus $\Delta\nu_-$ measures spacetime curvature in this case. 
Note that, assuming the frequency selectivity prescription for computing recoil boosts is true, this procedure is not possible with single-photon transitions, as detuning single-photon atom optics does not change the imparted recoil boost.
Lastly, if the phase shear modulation wavelength is comparable to the atom cloud size, in the two-photon case the measurement can be shifted to a spatial frequency more favorable for readout by introducing artificial shear (cf.\ \cite{sugarbaker2013enhanced,overstreet2018effective}), e.g.\ by frequency shifting the FBS.  This simply adds a known offset to the shear induced by spacetime curvature.

\section{Applications}
\subsection{Weak gravity and perturbative calculations}
\cref{eq:main} can facilitate explicit calculation of phase shear for realistic AI geometries.  To the best of our knowledge, there are no exactly solvable AI models in curved spacetime, so such calculations are necessarily approximate (whereas \cref{eq:main} makes no approximations with respect to atom and laser trajectories, and the refinement \cref{eq_freq_selectivity_result} is exact to the level of the semiclassical approximation).  In the common case that spacetime sectional curvature $K$ is nearly constant over the AI, such as for terrestrial AIs with curvature sourced by the Earth, the shear may be well approximated by simply multiplying the typical sectional curvature $K$ by the enclosed spacetime area $A$.  

For more refined calculations, a typical approach is to perturbatively solve for the AI arm trajectories, starting from trajectories in flat space \cite{GRAI}.  If we have solved for the AI arm trajectories to order $q$ in a metric perturbation, then integrating $K$ over the region bounded by these trajectories gives an expression for the shear which is accurate to order $q+1$, since $K$ is itself first order in the metric perturbation.  In particular, integrating $K$ over the unperturbed flat space atom trajectories gives a first order accurate estimate of the phase shear. 

Specializing to a static Newtonian metric of the form
\begin{equation}
\intd{s}^2 = \left(1+2\Phi(x)\right)\intd{t}^2 - \left(1-2\Phi(x)\right)\intd{x}^2,
\end{equation}
the sectional curvature is found to be $\Phi''(x)$ to leading order, reproducing a known result that phase shear measures gravity gradients \cite{roura2017circumventing,overstreet2018effective}.  In the non-relativistic limit we may thus express phase shear as 
\begin{multline}
\Delta p = -m \int_{t_0}^{t_f} \int_{x_\ell}^{x_u} \Phi''(x) \intd{x}\intd{t} \\= -m\int_{t_0}^{t_f}\Phi'(x_u)-\Phi'(x_\ell) \intd{t},
\end{multline}
where $t_0$ and $t_f$ are the times of the initial and final beamsplitters, and where $x_u(t)$ and $x_\ell(t)$ denote the trajectories of the upper and lower interferometer arms.  This expression recovers the non-relativistic interpretation of phase shear as the difference in impulse imparted to the upper and lower arms of the interferometer by the gravitational force. 

\subsection{Schwarzschild metric}

For the t-r plane of a Schwarzschild spacetime, the metric has the form 
\begin{equation}
\intd{s}^2 = \left(1 - \frac{r_s}{r}\right) \intd{t}^2 - \left(1-\frac{r_s}{r}\right)^{-1}\intd{r}^2
\end{equation}
with $r_s$ the Schwarzschild radius, and the sectional curvature is given by $K=-r_s/r^3$. At Earth's surface, $r_s = 8.9 \text{ mm}$ and $r = 6.4 \times 10^6\text{ m}$, so approximating $c^2K = -3100\text{ eotvos}$ as constant over the AI, the resulting phase shear wavelength is
\begin{equation}
\lambda = \frac{h}{m c |K A|} = 1.5 \text{ mm} \left(\frac{1\text{ m s}}{A/c}\right),
\label{eq:sch1}
\end{equation}
for $m=87 \text{ amu}$ (the approximate mass of $^{87}$Sr). This is readily observable for a few mm atom cloud size and a spacetime area $A/c \sim 1 \text{ m s}$, which is attainable in state-of-the-art 10-m drop tower experiments \cite{kovachy2015quantum}.  Practical measurement of phase shear due to spacetime curvature requires careful consideration of systematic effects contributing to longitudinal shear \cite{sugarbaker2013enhanced}.

\begin{figure}[t]
\centering
\includegraphics[width=0.8\linewidth]{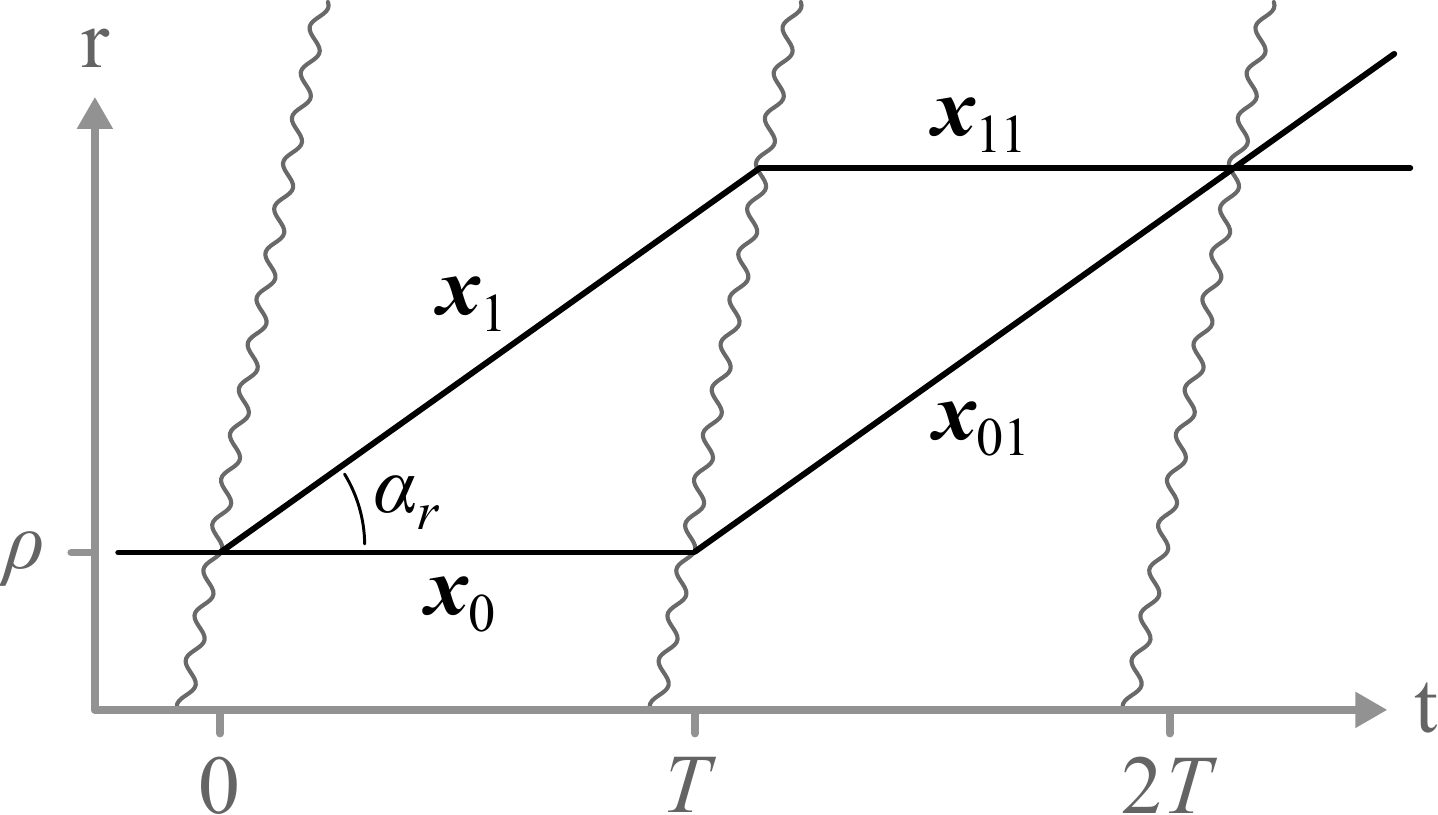}
\caption{\label{fig:schwarzschild} Unperturbed Mach-Zehnder atom trajectories for calculation of phase shear in the Schwarzschild spacetime.}
\end{figure}

For a more refined perturbative calculation, we suppose the atom starts with some initial spatial coordinate $r = \rho$ and take the unperturbed atom trajectories to be those of a single-photon Mach-Zehnder AI in Minkowski spacetime as shown in \cref{fig:schwarzschild}.  Letting $\alpha_r$ denote the recoil boost, the interferometer arm trajectories are given explicitly by
\begin{align*}
\begin{pmatrix} t_0(\tau) \\ r_0(\tau) \end{pmatrix} & = \begin{pmatrix} \tau \\ \rho \end{pmatrix} \qquad\qquad \qquad\qquad\qquad  \tau\in [0,T] \\
\begin{pmatrix} t_1(\tau) \\ r_1(\tau) \end{pmatrix} & = \begin{pmatrix} \tau \cosh(\alpha_r) \\ \rho + c\tau \sinh(\alpha_r) \end{pmatrix} \qquad\qquad  \tau \in [0,e^{\alpha_r}T]\\
\begin{pmatrix} t_{01}(\tau) \\ r_{01}(\tau) \end{pmatrix} & = \begin{pmatrix} T+ \left(\tau-T\right) \cosh(\alpha_r) \\ \rho + c\left(\tau-T\right) \sinh(\alpha_r) \end{pmatrix} \\
& \qquad\qquad\qquad\qquad \qquad\quad  \tau \in [T,(1+e^{\alpha_r})T] \\
\begin{pmatrix} t_{11}(\tau) \\ r_{11}(\tau) \end{pmatrix} & = \begin{pmatrix} e^{\alpha_r} T \cosh(\alpha_r) + \left(\tau-e^{\alpha_r}T\right) \\ \rho + e^{\alpha_r} c T \sinh(\alpha_r) \end{pmatrix} \\
& \qquad\qquad\qquad\qquad\qquad   \tau \in [e^{\alpha_r}T,(1+e^{\alpha_r})T].
\end{align*}
Integrating $K \intd{A} = \left(-c r_s/r^3\right)\intd{t}\intd{r}$ over the enclosed diamond-shaped region,
the resulting phase shear is given by 
\begin{multline}
\Delta p = -mc \int_{\rho}^{\rho + e^{\alpha_r} cT \sinh(\alpha_r)} \int_{\frac{r-\rho}{c\tanh\alpha_r}}^{\frac{r-\rho}{c\tanh\alpha_r}+T} \frac{-cr_s}{r^3} \intd{t}\intd{r} \\
= \frac{1}{2}mc^2Tr_s \left[\frac{1}{\rho^2} - \frac{1}{\left(\rho+e^{\alpha_r}cT\sinh(\alpha_r)\right)^2}\right].
\label{eq:sch-pert-shear}
\end{multline}
Again using terrestrial values $r_s = 8.9\text{ mm}$ and $\rho = 6.4 \times 10^6\text{ m}$ and assuming a $^{87}$Sr atom interferometer operating on the $^1S_0- {^3P_0}$ clock transition with mass $m=87\text{ amu}$, angular frequency $\omega = 2\pi c/(698\text{ nm})$, recoil boost $\alpha_r = \tanh^{-1}\left(\hbar\omega/(mc^2+\hbar\omega) \right) = 2.2 \times 10^{-11}$ (cf.\ \cref{eq:recoil}), and an enclosed spacetime area $A/c = e^{\alpha_r} \sinh(\alpha_r) c T^2 = 1 \text{ m s}$, we find the shear wavelength $\lambda \approx 1.5 \text{ mm}$, in agreement with the previous estimate \cref{eq:sch1}.

The sectional curvature of the Schwarzschild metric coincides to leading order with that derived from a Newtonian metric for a gravitating point mass.  Thus, to first order in the metric perturbation, the only deviations between the non-relativistic and relativistic calculations of phase shear arise from differences in the enclosed spacetime surface.  In \cref{eq:sch-pert-shear}, the only relativistic corrections are due to the finite speed of light, which causes the enclosed diamond to have slightly larger area than would be expected classically. 

\subsection{Other applications}
For long baseline terrestrial gradiometers, the \textit{differential} phase shear between the two AIs may be measurable.  For a vertical gradiometer at Earth's surface with baseline $1\text{ km}$ and spacetime area of $1000 \text{ m s}$ for each AI, such as proposed future versions of MAGIS \cite{MAGIS2021}, the difference in sectional curvature $\Delta K$ between the two AIs is $c^2\Delta K \approx 1.4\text{ eotvos}$. Thus if the shear of one AI is made zero as in \cite{roura2017circumventing}, the other will exhibit shear of wavelength $\approx 3.2 \text{ mm}$.  This may be a significant effect in future long baseline detectors. 

The gravitational influence of test masses \cite{overstreet2022observation} could be arranged to induce measurable phase shear.  The sectional curvature at the surface of a spherical test mass of density $\rho$ is $c^2 K = -\frac{8\pi}{3}G_N \rho$ (where $G_N$ is Newton's constant), or $|c^2 K|\sim 11000 \text{ eotvos}$ for tungsten.  This is several times larger than that due to terrestrial gravity at Earth's surface, but the sectional curvature of a test mass diminishes quickly with distance from the surface of the mass.  Thus in order to influence the phase shear of an interferometer, the AI must enclose a spacetime area \textit{near the test mass} of comparable magnitude to that required for terrestrial spacetime curvature to be observable.  Such a situation could be achieved via e.g.\ a co-moving test mass or a levitation sequence. 

Gravitational waves are not likely to produce measurable phase shear.  The sectional curvature of a gravitational wave is of order $c^{-2}\omega^2 h$, with $h$ the strain and $\omega$ the angular frequency.  For reasonable values $h \sim 10^{-21}$ and $\omega \sim 2\pi \times 1 \text{ Hz}$, $c^2 K \sim 3\times 10^{-11} \text{ eotvos}$, 14 orders of magnitude smaller than that at Earth's surface.

Lastly, we consider the effect of laser frequency noise.  An atom optic frequency offset $\Delta\omega$ alters the two-photon boost angle by $2\Delta\omega/m$ (cf.\ \cref{eq:FSSG}).  A laser frequency shift $\Delta\omega \sim 2\pi \times \left(1 \text{ kHz}\right)$ would induce spurious phase shear of wavelength $\pi c/\Delta\omega \sim 10^5 \text{ m}$, which is unobservable in practice. For an $N$ pulse large momentum transfer sequence with uncorrelated frequency shifts from pulse to pulse, this wavelength may be reduced by a factor of $\sim \sqrt{N}$, but is likely still unobservable for all proposed AI experiments, for which $N\sim 10^4$ is considered optimistic.

\section{Conclusion}

We have shown that phase shear of a collimated atom interferometer has a direct geometric interpretation as the integrated sectional curvature over the spacetime surface bounded by the interferometer arms and final beamsplitter null geodesic. For both single-photon and multi-photon atom interferometers phase shear thus provides a measurement of spacetime geometry which is independent of interferometer phase. This interpretation facilitates calculation of phase shear in non-trivial metrics. We have provided several examples where this effect is likely to be observable or already observed. The geometric interpretation of atom interferometer phase shear thus provides an elegant and powerful theoretical tool for understanding the connection between gravitation and atom interferometry.

\section{Acknowledgements}
This work was supported by the National Science Foundation QLCI Award No. OMA-2016244 and the Gordon and Betty Moore Foundation Grant GBMF7945.

\bibliography{bibs}

\appendix
\section{Lorentzian Angles Across Null Geodesics}
Our definition of Lorentzian angles between non-null vectors is equivalent to that of Helzer \cite{Helzer} or the real part only of the definition of Law \cite{Law}, and is equivalent to the definition of Birman and Nomizu \cite{BirmanNomizu} for time-like vectors. We will show that the Gauss-Bonnet formula as given by Law \cite{Law} can be extended to incorporate our definition of angles across null geodesics.  Helzer's notion of angle is more general still (and remarkably appeared well before all the alternative definitions of \cite{BirmanNomizu,Law,Dzan}), but we find the definition via parallel transport convenient for application to an AI final beamsplitter, where the ultimate goal is to compare two time-like momenta. 

\begin{figure}[t]
\centering
\includegraphics[width=0.95\linewidth]{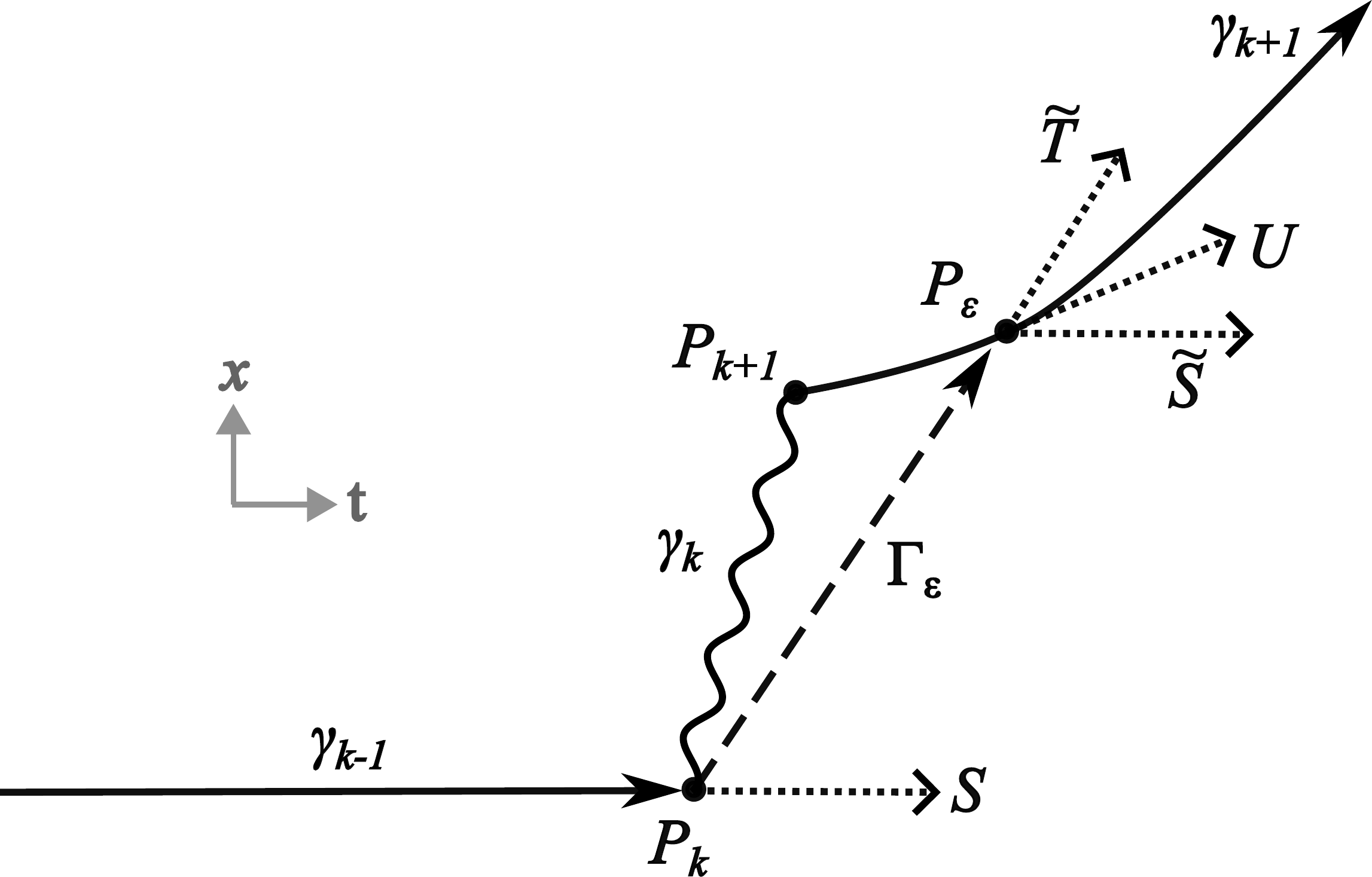}
\caption{\label{fig:null_approximation} Approximating a null geodesic segment $\gamma_k$ by a non-null geodesic segment $\Gamma_\varepsilon$.  See text for definitions.  Dotted lines indicate tangent vectors, solid lines are segments of the original path $\gamma$, and a dashed line is used for $\Gamma_\varepsilon$ only. For visual clarity, the tangent vector $T$ has been omitted, as it would overlap with $\Gamma_\varepsilon$. }
\end{figure}

Taking the real part only of the Gauss-Bonnet formula of \cite{Law} yields precisely our \cref{eq:GB}.  Suppose the boundary loop $\gamma$ consists of smooth segments $\gamma_j$, $j=0,\dots,n-1$. Let $P_j$ denote the common endpoint of $\gamma_{j-1}$ and $\gamma_j$ (with $j-1$ taken mod $n$) and let $\tau_j$ denote the corresponding parameter time such that $P_j = \gamma(\tau_j)$.  Suppose that exactly one segment $\gamma_k$ is a null geodesic and that it lies in a geodesically convex neighborhood.  Let the curve $\Gamma_\epsilon(\tau)$ be the unique geodesic defined on the interval $[\tau_k,\tau_{k+1}]$ starting at $P_k$ and terminating at $P_\epsilon := \gamma_{k+1}(\tau_{k+1}+\epsilon)$ (see \cref{fig:null_approximation}). 
For sufficiently small $\epsilon$, $\Gamma_\epsilon$ is non-null, for otherwise $\gamma_{k+1}(\tau_{k+1}+\epsilon)$ would lie on the light cone of $P_{k+1}$, which is impossible if $\gamma_{k+1}$ is non-null.  As $\epsilon\rightarrow 0$, $\Gamma_\epsilon(\tau)\rightarrow \gamma_{k}(\tau)$ for $\tau\in [\tau_k,\tau_{k+1}]$ (by the smoothness of the exponential map). 

Let $\alpha_k^{(\epsilon)}$ and $\alpha_{k+1}^{(\epsilon)}$ denote the angles at the intersections $\gamma_{k-1}$ with $\Gamma_\epsilon$ and $\Gamma_\epsilon$ with $\gamma_{k+1}$, respectively.  Let $S$ and $T$ be the tangents to $\gamma_{k-1}$ and $\Gamma_\epsilon$ at $P_k$, let $\widetilde S$, $\widetilde T$ be the parallel transport of $S,T$ along $\Gamma_\epsilon$ to $P_\epsilon$, and let $U$ be the tangent to $\gamma_{k+1}$ at $P_\epsilon$.  Note that $\widetilde T$ is the tangent to $\Gamma_\epsilon$ at $\Gamma_\epsilon(\tau_{k+1})$ since $\Gamma_\epsilon$ is geodesic.  Parallel transport preserves angles (since it preserves inner products), and thus $\alpha_k^{(\epsilon)}$ is equal to the angle from $\widetilde S$ to $\widetilde T$.  By additivity of angles, $\alpha_k^{(\epsilon)}+\alpha_{k+1}^{(\epsilon)}$ equals the angle from $\widetilde S$ to $U$.  By continuity of parallel transport with respect to initial conditions, as $\epsilon\rightarrow 0$, $\alpha_k^{(\epsilon)}+\alpha_{k+1}^{(\epsilon)}$ converges to the angle across the null geodesic $\gamma_k$. 

Apply the Gauss-Bonnet formula of \cite{Law} to the modified curve $\gamma_\epsilon$ which follows $\gamma$ until reaching $P_k$, then follows $\Gamma_\epsilon$ to $P_{\epsilon}$, then resumes following $\gamma$ for all remaining time.  By the remarks of the above paragraph, the sum of angles of $\gamma_\epsilon$ converges to that of $\gamma$.  Since $\gamma_k$ and $\Gamma_\epsilon$ are both geodesic, the geodesic curvature of $\gamma_\epsilon$ differs from that of $\gamma$ only on a short segment of $\gamma_{k+1}$, so that $\oint_{\gamma_\epsilon} k_g \intd{s}$ converges to $\oint_{\gamma} k_g \intd{s}$ also.  Lastly, the sectional curvature integral for $\gamma_\epsilon$ differs from that of $\gamma$ only by a wedge shaped region with vanishing area as $\epsilon\rightarrow 0$.  Thus each term in the Gauss-Bonnet formula for $\gamma_\epsilon$ separately converges to that with our definition of angle across a null geodesic, proving our generalized Gauss-Bonnet theorem in this case.

If the boundary curve $\gamma$ has multiple null geodesics, inductively applying the same construction proves that the Gauss-Bonnet formula holds in this case as well. 

We make two remarks on subtleties of applying Gauss-Bonnet in special cases: First, any spacetime surface bounded by the AI may be used to compute the sectional curvature integral.  If the surface is deformed in the ambient 3+1 dimensional spacetime, the sectional curvature witnessed at particular points will of course vary, but the integrated curvature must remain fixed. This allows some flexibility in choosing convenient surfaces for calculation. Second, in dimensions higher than 1+1, it may happen that the same atom optics null geodesics do not address both arms of the AI. In this case, the Gauss-Bonnet formula may be applied to a loop formed by the AI arms, two null geodesic segments associated with the trajectories of FBS photons which address the two AI arms, and a short spacelike segment connecting these two null segments. 

\end{document}